\documentclass{iaus}
\usepackage{graphicx}

\title[JD 11.~~Spatial Evolution of Stellar Structures] 
{The Spatial Evolution of Stellar Structures in the LMC/SMC}

\author[Bastian, Gieles, Ercolano, \& Gutermuth]   
{Nate Bastian$^1$, Mark Gieles$^2$, Barbara Ercolano$^1$
 \and Robert Gutermuth$^3$}

\affiliation{$^1$Institute of Astronomy, University of Cambridge, Madingley Road, Cambridge, CB3 0HA, UK\\ email: {\tt bastian@ast.cam.ac.uk; be@ast.cam.ac.uk} \\[\affilskip]
$^2$European Southern Observatory, Casilla 19001, Santiago 19, Chile  \\email: {\tt mgieles@eso.org}\\[\affilskip]
$^3$Harvard-Smithsonian Center for Astrophysics, 60 Garden Street, Cambridge, MA 02138, USA  \\email: {\tt rgutermuth@cfa.harvard.edu}}

\pubyear{2008}
\volume{256}  
\pagerange{119--126}
\jname{The Magellanic System: Stars, Gas, and Galaxies}
\editors{Jacco Th. van Loon \& Joana M. Oliveira, eds.}
\begin{document}

\maketitle

\begin{abstract}
We present an analysis of the spatial distribution of various stellar populations within the Large and Small  Magellanic Clouds.  We use optically selected stellar samples with mean ages between $\sim9$ and $\sim1000$~Myr, and existing stellar cluster catalogues to investigate how stellar structures form and evolve within the LMC/SMC.  We use two statistical techniques to study the evolution of structure within these galaxies, the $Q$-parameter and the two-point correlation function (TPCF).  In both galaxies we find the stars are born with a high degree of substructure (i.e. are highly fractal) and that the stellar distribution approaches that of the 'background' population on timescales similar to the crossing times of the galaxy ($\sim80$~Myr \& $\sim150$~Myr for the SMC/LMC respectively).  By comparing our observations to simple models of structural evolution we find that 'popping star clusters' do not significantly influence structural evolution in these galaxies.  Instead we argue that general galactic dynamics are the main drivers, and that substructure will be erased in approximately the crossing time, regardless of spatial scale, from small clusters to whole galaxies.  This can explain why many young Galactic clusters have high degrees of substructure, while others are smooth and centrally concentrated.  We conclude with a general discussion on cluster 'infant mortality',  in an attempt to clarify the time/spatial scales involved. 
\end{abstract}

\firstsection 
\section{Introduction}

Most, if not all, stars are thought to be born in a {\it clustered} or fractal distribution, which is usually interpreted as being due to the imprint of the gas hierarchy from which stars form (e.g. Elmegreen \& Efremov~1996, Elmegreen et al.~2006, Bastian et al.~2007).  Older stellar distributions, however, appear to be much more smoothly distributed, begging the questions; 1) what is the main driver of this evolution? and 2) what is the timescale for the natal structure to be erased?

It has been noted that many nearby star forming clusters also have  
hierarchical structure seemingly dictated by the structure of the  
dense gas of natal molecular clouds (e.g. Lada \& Lada 2003).  Using  
statistical techniques, Gutermuth et al. (2005) studied three  
clusters of varying degrees of embeddedness and demonstrated that the  
least embedded cluster was also the least dense and the least  
substructured of the three. That result hinted at the idea that the  
youngest clusters are substructured, but that dynamical interactions  
and ejection of the structured gas contributes to the evolution and  
eventual erasure of that substructure in approximately the cluster  
formation timescale of a few Myr (Palla \& Stahler 2000). By having an  
accurate model of the spatial evolution of stellar structures we can  
approach a number of fundamental questions about star-formation,  
including what is the percentage of stars born in "clusters" and  
whether this depends on environmental conditions. Using automated  
algorithms on infrared Spitzer surveys of star-forming sites within  
the Galaxy (e.g. Allen et al. 2007), such constraints are now becoming  
possible.

In this contribution we present results of two recent studies on the evolution of structure in the LMC (Bastian et al.~2008) and SMC (Gieles et al.~2008).

\begin{figure}[h]
\begin{center}
 \includegraphics[width=3.4in]{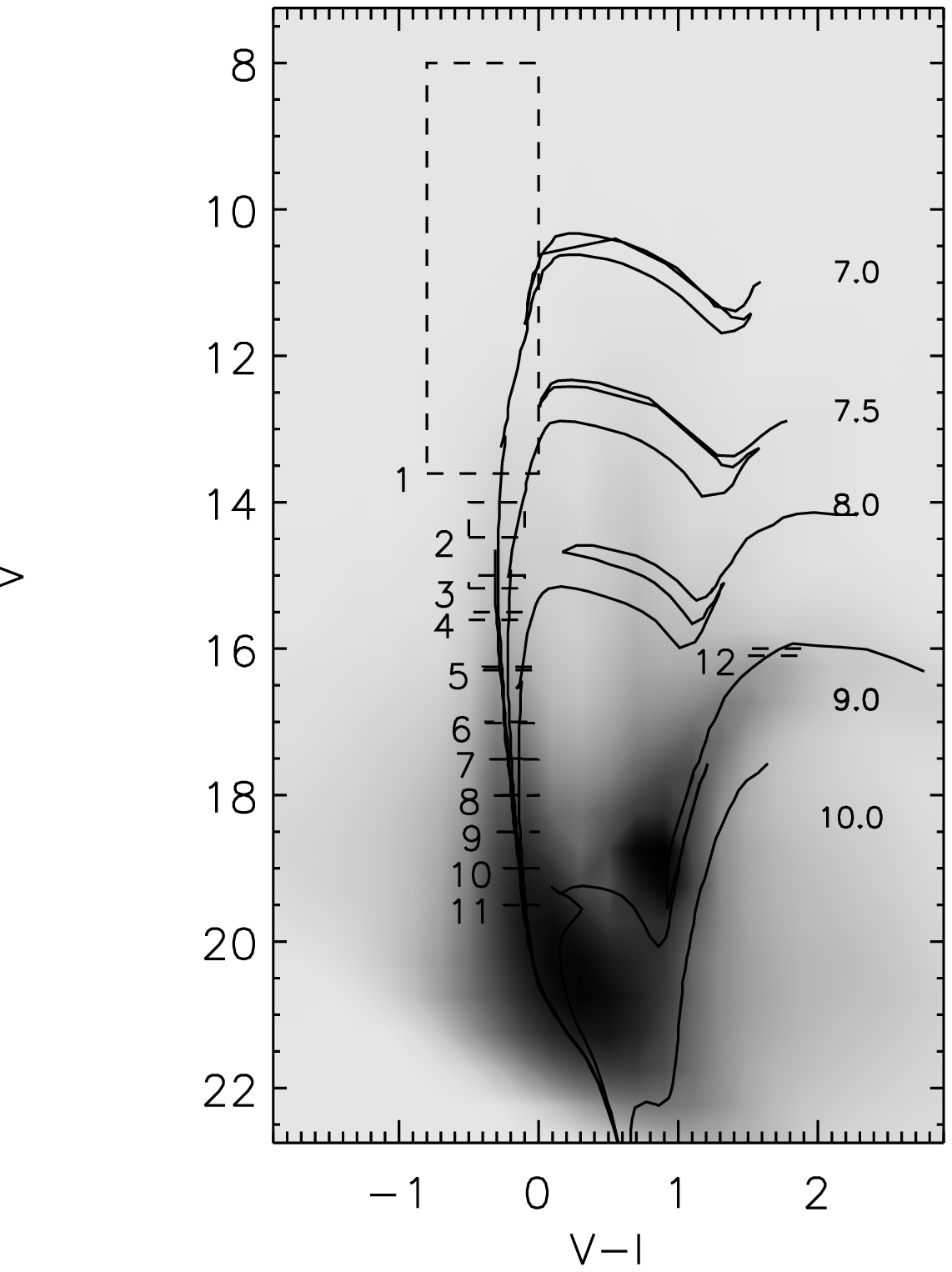} 
 \includegraphics[width=5.4in]{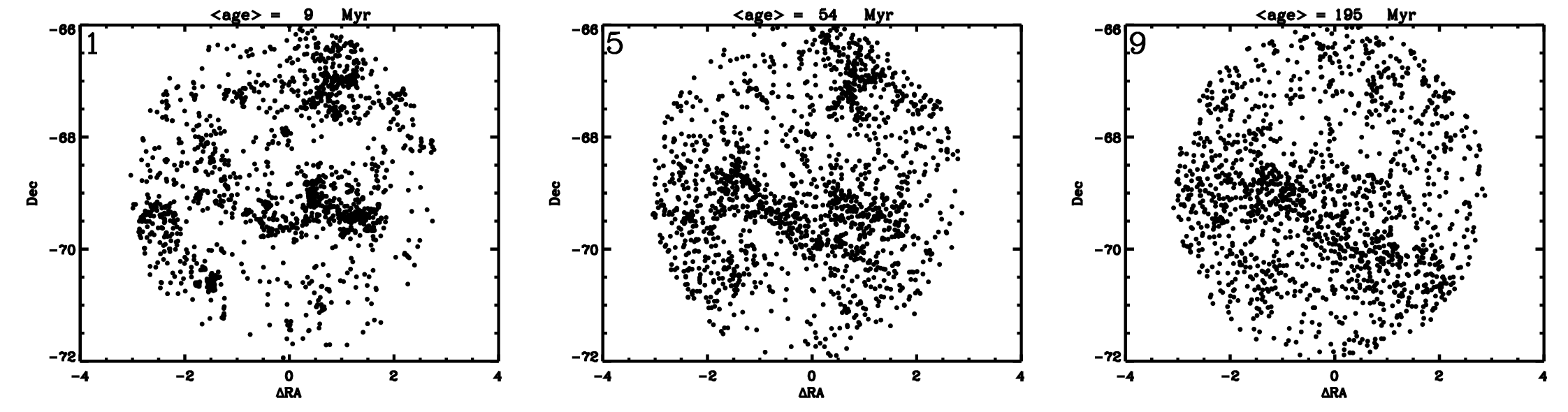} 
 \caption{{\bf Top:} A V-I vs. V CMD of stars in the LMC.  The selected boxes are shown, along with theoretical stellar evolutionary isochrones. {\bf Bottom:} The spatial distribution of stars in boxes 1, 5, \& 9 in the LMC.  The mean age of the stellar distributions is given at the top of each panel.}
   \label{fig:cmd}
\end{center}
\end{figure}

\section{Datasets and Techniques}

Our main dataset is the Magellanic Cloud Photometic survey (Zaritsky et al.~2004), consisting of several million stars with UBVI photometry in the LMC/SMC.  By using colour and magnitude cuts 
in the colour-magnitude diagrams (CMDs) of the LMC/SMC stars, we can select populations of different mean ages.  An example CMD (where the grey scale is the logarithm of the density of stars at that colour/magnitude) for the LMC is shown in top panel of Fig.~\ref{fig:cmd}, where the bottom panels show the spatial positions of stars in a sample of age boxes.    The mean ages of these boxes are derived using Monte Carlo sampling, assuming a Salpeter IMF, of theoretical stellar isochrones (Girardi et al. 2002).   Additionally, for comparison, we use the ages and positions of stellar clusters from the Hunter et al.~(2003) catalogue.

In order to study the evolution of structure, we employ two statistical methods.  The first is the $Q$-parameter (Cartwright \& Whitworth~2004) which is based on minimum spanning trees (MST)\footnote{An MST is formed by connecting all points (spatial positions in this case) in order to form a unified network, such that the sum of all of the connections, known as 'edges' or 'branches', is minimized, and no closed loops are formed.}, while the second is the two-point correlation function.  

The $Q$-parameter uses the normalised mean MST branch length, $\overline{\rm m}$, and the normalised distance between all sources within a region, $\overline{\rm s}$.  The ratio between these to quantities, $\overline{\rm m}$/$\overline{\rm s}$ = $Q$, is able to distinguish between a power-law (centrally concentrated) profile and a profile with sub-structure (i.e. a fractal distribution).  Additionally, this parameter can quantify the index of the power-law or the degree of sub-substructure, which, if one assumes is due to a fractal nature, its fractal dimension can also be estimated.  Assuming a three dimensional structure, if Q is less than 0.79 then the region is fractal, larger than 0.79 refers to a power-law structure, and a value of 0.79 implies a random distribution (a 3D fractal of dimension 3 is a random distribution).  However, if the distribution is two dimensional, as is approximately true if one is looking at a disk-like galaxy face on (like the LMC), then a random distribution has a Q value of 0.72.

The two-point correlation function (TPCF) determines the distance between all possible pairs of stars, shown as a histogram, which is then normalized to that of a reference distribution, i.e. $N_{\rm links}$ / $N_{\rm reference}$ where $N_{\rm reference}$ is taken as a smooth, centrally concentrated power-law distribution as observed in old stellar populations for each galaxy.   This is similar to that done by Gomez et al.~(1993) who used the TPCF to study the distribution of pre-main sequence stars in Taurus.  We then measure the slope and zero-point of the resulting distributions.

\begin{figure}[h]
\begin{center}
  \includegraphics[width=3.4in]{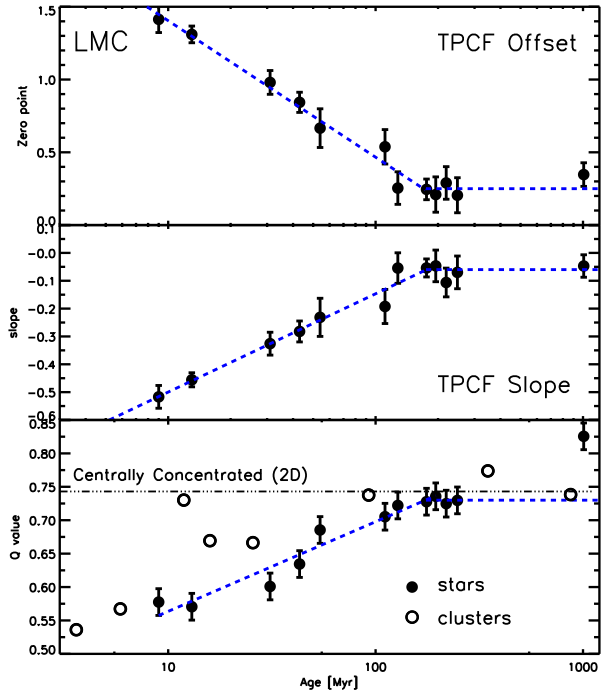}  
 \caption{Results for the LMC.  {\bf Top:} The $Q$-parameter for each of the age boxes.  {\bf Bottom:} The evolution of the slope and zero-point of the two-point correlation function (TPCF).}
   \label{fig:results-lmc}
\end{center}
\end{figure}

\begin{figure}[h]
\begin{center}
 \includegraphics[width=3.4in]{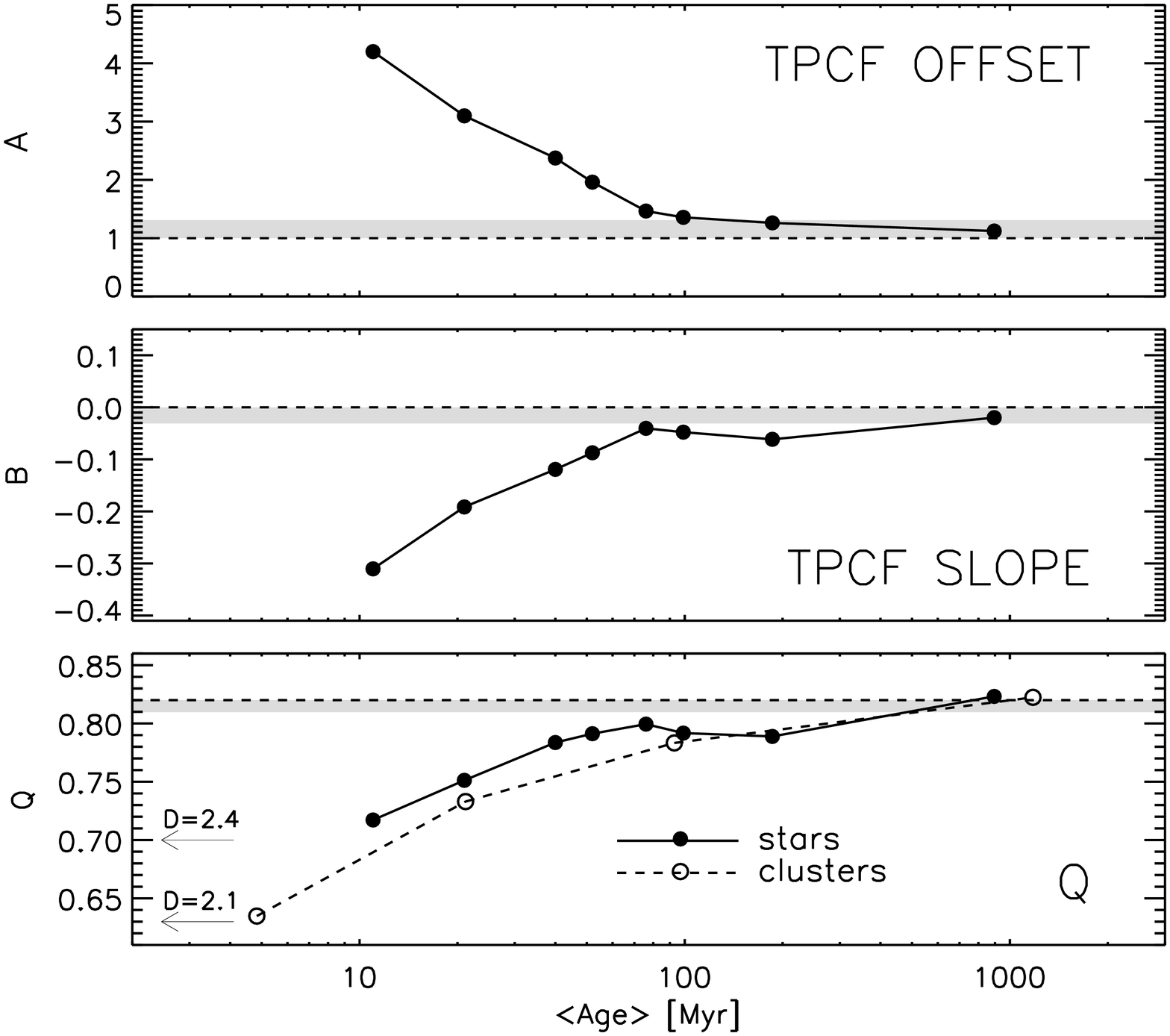} 
 \caption{Results for the SMC.  {\bf Top and Middle:} The evolution of the slope and zero-point of the two-point correlation function (TPCF).  {\bf Bottom:} The $Q$-parameter for each of the age boxes. }
   \label{fig:results-smc}
\end{center}
\end{figure}

The results from these two methods are given in Fig.~\ref{fig:results-lmc} for the LMC and Fig.~\ref{fig:results-smc} for the SMC.  For the LMC we see that both the $Q$-parameter method and both measurements from the TPCF method, show that substructure is erased (i.e. where the distributions become flat) on a timescale of $~\sim175$~Myr (see Bastian et al.~2008 for details).  For the SMC, substructure appears to be removed in $\sim80$~Myr (see Gieles et al.~2008 for details).

\section{Implications}

Two scenarios have been put forward to explain the rapid evolution of substructure in galaxies.  The first scenario, 'popping star clusters' (Kroupa~2002) is based on the idea infant mortality, or early dissolution of star cluters.  When a cluster forms, a large fraction of the gas of the natal cloud or core remains after star-formation has terminated (i.e. there is a non-100\% star-formation efficiency).  This gas may be removed explosively due to the energy input of the most massive stars in the clusters, thus removing a substantial fraction of the gravitational potential of the cluster.  This leaves the stars in a super-virial state, and depending on the star-formation efficiency, the cluster may expand (approximately at its velocity dispersion), or 'pop', spreading its stars into the field.  The vast majority of clusters in the LMC/SMC are low mass systems, and hence are expected to have small velocity dispersions, on the order of $< 1$~km/s.  With these small velocities, it would take $\sim1$~Gyr for a star in either galaxy to move one galactic radius.  Additionally, in this scenario we would expect to see differences in the evolution of the stars vs. that of the clusters, as the clusters are not affected by this, only by the general dynamics of the galaxy.  However, for both the LMC/SMC we see that the $Q$-parameter of the clusters and stars have a very similar evolution (see Bastian et al.~2008 for detailed simulations of this effect).

Hence we are left to conclude that general galactic dynamics is the main driver of the removal of substructure within galaxies.  This does not imply that infant mortality does not exist within these galaxies, only that it does not have a significant impact on the evolution of substructure within the LMC/SMC.  If general dynamics are the main driver, then we expect that it should be independent of spatial scale, i.e. always removed on approximately the crossing time.  This can be readily seen in many young clusters in the Galaxy as well as in the LMC/SMC, where the central regions of clusters are smooth and show little or no substructure (i.e. where the crossing time is short) whereas in the outer regions (where the crossing time is significantly longer) substructure remains for longer periods.  With this hypothesis in hand, it is relatively easy to calculate where substructure should be present on any scale, if the age, size and velocity dispersion (or crossing time) is known.

\subsection{Clustered star formation vs. star formation in clusters}

In Bastian et al.~(2007) we showed that most, if not all, OB stars in M33 are formed in a clustered fashion, i.e. part of a hierarchical or fractal distribution down to the resolution limit of the sample.  However, this is not to say that all stars form in dense (e.g. $R_{\rm eff} \sim 3$~pc) clusters, in fact no characteristic size of the star-forming regions was found.  Only a small fraction of star-formation takes place in clusters which will survive long enough (i.e. $>3$~Myr) and be dense enough, to be selected in optical samples (Lada \& Lada~2003, Gieles  \& Bastian~2008, Bastian~2008), on the order of a few percent.  This follows naturally from a fractal gas distribution and a threshold density for star formation along with a threshold efficiency for bound cluster formation (Elmegreen~2008).

Since young clusters do not necessarily look like older, evolved, and relaxed clusters (i.e. young embedded clusters are often extremely hierarchical) it is difficult to define which stars belong to a young 'cluster', as the 'clusters' themselves are somewhat arbitrarily defined.  Hence the term 'infant mortality' itself is fairly ill-defined, as one must pre-select the physical scale of interest, e.g. structures which are $\sim1$~pc (Lada \& Lada~2003), or $\sim100$~pc~(Pellerin et al. 2007) in scale.  Additionally, 'infant mortality' was originally defined as the disruption of stellar structures due to the removal of the left over gas (i.e. non-100\% star-formation), which implies that the gas + stars were originally bound.  However, it is arguable whether structures larger than a few pc fit this criterion, and hence their dissolution should not be confused with 'infant mortality', but instead simply represents an {\it initially} unbound system which is being pulled apart by general galactic dynamics.  Finally, the effects due to gas expulsion are largely over by a few tens of Myr (e.g. Bastian \& Goodwin~2006), hence cluster disruption beyond this age range requires altogether different mechanisms, and hence should be treated separately (e.g. Gieles, Lamers, \& Portegies Zwart~2007).

\end{document}